\documentclass[fleqn,usenatbib]{mnras}

\usepackage{newtxtext,newtxmath}
\usepackage[T1]{fontenc}

\DeclareRobustCommand{\VAN}[3]{#2}
\let\VANthebibliography\thebibliography
\def\thebibliography{\DeclareRobustCommand{\VAN}[3]{##3}\VANthebibliography}

\usepackage{graphicx}	% Including figure files
\usepackage{amsmath}	% Advanced maths commands
\usepackage{hyperref}
\usepackage{cprotect}
\usepackage{multirow}

\newcommand{\jr}{\color{black}}

\title[Atmospheric composition from escaping H/He]{Using observations of escaping H/He to constrain the atmospheric composition of sub-Neptunes}

\author[Rogers,  Owen,  Schreyer \& Kirk]{
James G. Rogers$^{1}$\thanks{E-mail: jr2011@cam.ac.uk}, James E. Owen$^{2}$, Ethan Schreyer$^{3}$ \& James Kirk$^{2}$
\\
$^{1}$Institute of Astronomy, University of Cambridge, Madingley Road, Cambridge CB3 0HA, United Kingdom\\
$^{2}$Astrophysics Group, Department of Physics, Imperial College London, Prince Consort Rd, London, SW7 2AZ, UK\\
$^{3}$Department of Astronomy and Astrophysics, University of California, Santa Cruz, CA 95064, USA\\
}

\date{Accepted XXX. Received YYY; in original form ZZZ}

\pubyear{\the\year{}}

\begin{document}
\label{firstpage}
\pagerange{\pageref{firstpage}--\pageref{lastpage}}
\maketitle

\begin{abstract}
The internal composition of sub-Neptunes remains a prominent unresolved question in exoplanetary science. We present a technique to place constraints on envelope mean molecular weight that utilises observations of escaping hydrogen or helium exospheres. This method is based on a simple timescale argument, which states that sub-Neptunes require a sufficiently large hydrogen or helium reservoir to explain on-going escape at their observed rates. This then naturally leads to an upper limit on atmospheric mean molecular weight. {\jr{We formalise this argument within a Bayesian inference model and apply it}} to the archetypal sub-Neptunes GJ-436 b, TOI-776 b and TOI-776 c, which have all been observed to be losing significant hydrogen content as well as relatively featureless transit spectra when observed with \textit{JWST}. Combining constraints from atmospheric escape and transit spectroscopy in the case of TOI-776 c allows us to tentatively rule out the high mean molecular weight scenario, pointing towards a low mean molecular weight atmosphere with high-altitude aerosols muting spectral features in the infra-red. Finally, we reframe our analysis to the hycean candidate K2-18 b, which has also been shown to host a tentative escaping hydrogen exosphere. If such a detection is robust, we infer a hydrogen-rich envelope mass fraction of $\log_{10} f_\text{env} = -1.67\pm0.78$, which is inconsistent with the hycean scenario at the $\sim 4\sigma$ level. This latter result requires further observational follow-up to confirm.
\end{abstract}


%%%%%%%%%%%%%%%%%%%%%%%%%%%%%%%%%%%%%%%%%%%%%%%%%%

%%%%%%%%%%%%%%%%% BODY OF PAPER %%%%%%%%%%%%%%%%%%

\section{Introduction} \label{sec:intro}
Recent characterisation of sub-Neptune exoplanets has brought about an apparent contradiction. On the one hand, it is widely believed that the existence of the ``radius valley'', a paucity in planet occurrence at $\sim 1.8R_\oplus$, is a product of atmospheric escape \citep[e.g.][]{Owen2013,LopezFortney2013,Fulton2017,VanEylen2018}. A subset of the primordial population of sub-Neptunes, being silicate-rich interiors hosting hydrogen-dominated envelopes, can have their envelopes stripped as a result of stellar irradiation, revealing naked planetary cores and forming the population of super-Earths below the radius valley \citep[e.g.][]{Owen2017}. This argument has been supported by multiple studies, including direct observations of escaping hydrogen or helium exospheres \citep[e.g.][]{DosSantos2023a} as well as demographic analyses of the exoplanet radius distribution as a function of orbital period \citep{Gupta2019,Rogers2021}, stellar mass \citep{Gupta2020,Rogers2021b,Petigura2022,Ho2024} and stellar age \citep{Fernandes2022,Vach2024a,Fernandes2025,Rogers2025b}.

One of the clear expectations of this theoretical framework was the existence of low mean molecular weight sub-Neptune atmospheres when characterised with spectroscopy from instruments such as the Hubble Space Telescope (\textit{HST}) or the James Webb Space Telescope (\textit{JWST}). This has been shown to be true for some temperate sub-Neptunes, such as TOI-270 d and K2-18 b \citep[e.g.][]{Madhusudhan2023,Benneke2024,Holmberg2024}. In these cases, the low mean molecular weights produce puffy atmospheres with large scale heights, allowing for deep molecular absorption features to be detected. This is also true for the hot sub-Neptune TOI-421 b which exhibits large spectral features \citep{Davenport2025}. However, for many sub-Neptunes, atmospheric spectra have been observed to be featureless, such GJ-436 b \citep{Mukherjee2025}, GJ 3090 b \citep{Ahrer2025}, and those targeted in the COMPASS \textit{JWST} program including TOI-776 b, TOI-776 c and TOI-836 c \citep{Wallack2024,Alderson2024,Alderson2025,Teske2025} which have all returned relatively flat spectra. The contradiction exists, however, because all of the aforementioned planets host observed escaping hydrogen or helium exospheres \citep{Ehrenreich2015,Zhang2025,Loyd2025,Ahrer2025}.

The specific contradiction is thus: if several sub-Neptune atmospheres {\jr{with flat spectra}} are consistent with having a high mean molecular weight and therefore host relatively low abundances of hydrogen and helium, then how are they also observed to be losing significant quantities of hydrogen or helium via atmospheric escape? One likely possibility is that their atmospheres are hydrogen/helium-rich but host high altitude aerosols that introduce broadband grey opacity that mute the spectral features \citep[e.g.][]{Gao2021,Lodge2024,Moran2025,Owen2025}. However, several \textit{JWST} studies have shown that a degeneracy exists between the presence of low pressure aerosols and high mean molecular weight atmospheres \citep[e.g.][]{Wallack2024,Alderson2024,Alderson2025,Teske2025}.

In this paper, we introduce a simple analytic timescale argument that can aid in constraining the mean molecular weight of a sub-Neptune atmosphere based on observations of escaping hydrogen or helium. We show that such an observation naturally places an upper limit on the envelope's mean molecular weight and can be combined with constraints from atmospheric spectroscopy to further constrain atmospheric properties including composition and opaque aerosol pressure level.

\section{Part I: Constraints from Mass Loss} \label{sec:partI}

\subsection{A sketch of the idea} \label{sec:partI_sketch}
Consider a sub-Neptune with core mass of $M_\text{c} = 5 M_\oplus$ and envelope mass fraction of $f_\text{env} = M_\text{env} / M_\text{c} = 1\%$. Let us assume the envelope only consists of hydrogen, H$_2$, with mass fraction $X$, helium, He, with mass fraction $Y$, and ``metals'', with mass fraction $Z$. By definition, $X+Y+Z=1$. Now, for a typical observed hydrogen mass loss rate of $\dot{M}_\text{H}=10^9 \text{ g s}^{-1}$, the hydrogen mass loss timescale, $\tau_{\dot{M}_\text{H}}$, is given by:
\begin{equation} \label{eq:MassLossTimescale}
    \tau_{\dot{M}_\text{H}} = \frac{M_\text{c} f_\text{env} X}{\dot{M}_\text{H}},
\end{equation}
where the product of $M_\text{c} f_\text{env} X$ is equal to the total hydrogen mass in the planet's envelope. An observed mass loss rate implies that the planet's hydrogen reservoir must have been sufficiently large to sustain this mass loss rate over \textit{at least} the planet's lifetime. In other words, if the hydrogen reservoir were not large enough, then we could not have observed hydrogen mass loss today. Thus, the minimum hydrogen mass the planet could host today is the result of the planet losing mass at the observed rate for its lifetime. Assuming the planet's age is equal to the host star's age, $t_*$, we can thus rewrite Equation \ref{eq:MassLossTimescale} as an inequality:
\begin{equation} \label{eq:X-inequality}
    X \geq \frac{\dot{M}_\text{H} t_*}{M_\text{c}f_\text{env}}.
\end{equation}
In reality, as a sub-Neptune evolves, one expects its hydrogen mass loss rate to substantially decrease over time due to a combination of lower XUV activity from its host star, and its own planetary contraction which reduces its cross section to incoming stellar irradiation. As a result, the expected hydrogen mass fraction may be significantly larger than this limit. As such, these limits are both conservative and robust. {\jr{We also highlight that the construction of our timescale argument is independent of assumptions on atmospheric escape physics, such as the dominant heating mechanism \citep[e.g.][]{Owen2024,Misener2026} or metal-dependent escape efficiencies \citep[e.g.][]{Yoshida2022,Zhang2025,Broome2025,Yoshida2025}.}}

For our hypothetical sub-Neptune with an assumed age of $t_* = 5$~Gyr, we find a lower limit on hydrogen mass fraction of $X \geq 0.53$. This limit can be recast as an upper limit on envelope mean molecular weight, which is given as:
\begin{equation} \label{eq:mmw-inequality}
    \mu \leq \bigg( \frac{X}{\mu_{\text{H}_2}} + \frac{Y}{\mu_{\text{He}}} + \frac{Z}{\mu_Z} \bigg)^{-1},
\end{equation}
where $\mu_{\text{H}_2}=2.016$~g mol$^{-1}$, $\mu_{\text{He}}=4.003$~g mol$^{-1}$ are the relevant molecular weights for hydrogen and helium. For simplicity, we can assume that the averaged molecular weight of all metal species is equal to that of H$_2$O, $\mu_Z=18.015$~g mol$^{-1}$, {\jr{which is justified since oxygen is the dominant heavy element at Solar abundances and predominantly exists as H$_2$O in chemical equilibrium.}} Finally, we also assume a Solar H$_2$/He mass ratio of $X/Y = 0.7491 / 0.2377$ \citep[e.g.][]{Lodders2003}, which leads to an upper limit on envelope mean molecular weight of $\mu \leq 3.09 \text{ g mol}^{-1}$. 

\begin{figure}
	\includegraphics[width=\columnwidth]{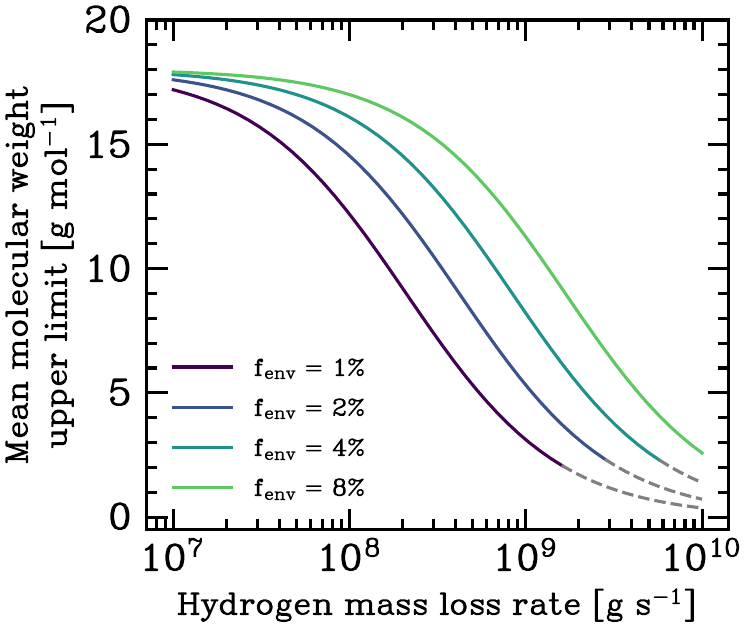}
    \centering
        \cprotect\caption{The upper limit on envelope mean molecular weight is shown as a function of observed hydrogen mass loss rate for various envelope mass fractions. This applies to a $5$~Gyr old sub-Neptune with a core mass of $5$~M$_\oplus$ and assuming a Solar hydrogen-to-helium mass fraction following Equations \ref{eq:X-inequality} and \ref{eq:mmw-inequality}. Dashed grey lines denote upper limits on mean molecular weight that are below Solar values. The higher the observed mass loss rate and the smaller the envelope mass, the lower the mean molecular weight needs be in order to explain the escaping hydrogen observations.} \label{fig:mlr-mmw-relation} 
\end{figure} 

Figure \ref{fig:mlr-mmw-relation} demonstrates the relation between observed hydrogen mass loss rate and the inferred upper limit on mean molecular weight (via Equations \ref{eq:X-inequality} and \ref{eq:mmw-inequality}). As mass loss rates increase, the hydrogen reservoir must necessarily increase to sustain this mass loss, which in turn leads to a decreasing upper limit on mean molecular weight. This relation is shown for various envelope mass fractions, demonstrating how increasing the envelope mass allows for a lower hydrogen mass fraction to still explain the observed mass loss. This subsequently weakens the constraint on mean molecular weight.

\subsection{Inference model for escaping hydrogen} \label{sec:partI_inference_model}

The previous sketch demonstrates how an observation of on-going atmospheric escape can be leveraged to place upper limits on a sub-Neptune's envelope mean molecular weight. However, several assumptions were made, including values for the planet's core mass and envelope mass fraction, which are not \textit{a priori} known for an observed exoplanet and must also be inferred. We have also neglected measurement uncertainty in observed planet properties which must be accounted for. Finally, we have ignored other measured observables, such as planet mass and radius, which help constrain planet properties. We now present a simple Bayesian inference model with the goal of placing statistical upper limits on envelope mean molecular weight for a given sub-Neptune. 

\begin{table*} 
\begin{center} 
\begin{tabular}{l l l l} 
 \hline
 Planet name & TOI-776 b & TOI-776 c & GJ 436 b\\ [0.5ex] 
 \hline\hline
 Radius $[R_\oplus]$ & $1.798\pm0.078^a$ & $2.047\pm0.081^a$ & $4.170 \pm 0.168^c$\\ 
 \hline
Mass $[M_\oplus]$ & $5.0 \pm 1.6^a$ & $6.9 \pm 2.6^a$ & $22.1 \pm 2.3^c$ \\
 \hline
 Semi-major axis [AU] & $0.0653\pm0.0015^a$ & $0.1001\pm0.0024^a$ & $0.0291 \pm 0.0015^d$\\
 \hline
 Equilibrium temperature [K] & $520\pm12^a$ & $415\pm14^a$ & $686\pm10^d$\\
 \hline
 Stellar mass [$M_\odot$] & $0.542\pm0.039^a$ & $0.542\pm0.039^a$ & $0.47\pm0.07^c$\\ [1ex] 
 \hline
 Stellar age [Gyr] & $1-4^b$ & $1-4^b$ & $1-6^e$\\ [1ex] 
 \hline
 $\log_{10}$(mass loss rate [g s$^{-1}$]) & $9.25^{+0.63\,b}_{-0.86}$ & $9.18^{+0.56\,b}_{-0.65}$ & $9.40^{+0.38\,f}_{-0.40}$\\
 \hline
\end{tabular}
\caption{Planetary and stellar parameters for our test-case planets. (a) \citet{Fridlund2024}; (b) \citet{Loyd2025}; (c) \citet{Maciejewski2014}; (d) \citet{Turner2016}; (e) \citet{Torres2008}; (f) \citet{Schreyer2024}.}
\label{tab:planets}
\end{center}
\end{table*} 

We assume that our planet of interest has an observed mass, radius and hydrogen mass loss rate. We assume Gaussian measurement uncertainties for mass and radius e.g. $R_\text{p,obs}\pm\sigma_{\text{R}_\text{p}}$, $M_\text{p,obs}\pm\sigma_{\text{M}_\text{p}}$, where $\sigma$ represents the relevant uncertainty as shown in Table \ref{tab:planets}.\footnote{The assumption of Gaussian measurement uncertainties can be removed by incorporating the full posteriors for each planet property within the inference analysis.} Our model parameters are the planet's core mass, $M_\text{c}$, core radius, $R_\text{c}$, envelope mass fraction, $f_\text{env}$, envelope mean molecular weight, $\mu$, hydrogen mass loss rate, $\dot{M}_\text{H}$, semi-major axis, $a$, equilibrium temperature, $T_\text{eq}$, host stellar mass, $M_*$, and host stellar age, $t_*$. Other properties of the planet are derived from these variables, such as planet mass, $M_\text{p} = M_\text{c} (1 + f_\text{env})$. We continue to assume a Solar H$_2$/He mass ratio of $X/Y = 0.7491 / 0.2377$ in the atmosphere \citep{Lodders2003}. In reality, if helium is inefficiently dragged in the hydrogen outflow, then mass fractionation could act to reduce this ratio during atmospheric escape \citep[e.g.][]{Zahnle1986}. However, in our case of observed escaping hydrogen, fractionation would in fact strengthen our compositional constraint due to helium's larger mean molecular weight.\footnote{Furthermore, since our adopted Solar H$_2$/He mass ratio is close to Big Bang nucleosynthesis values, one would not expect this ratio to increase substantially.} 

We assume the atmospheric ``metal'' mass fraction, $Z$, to have the same molecular weight as water vapour, but can in fact represent a mixture of different high mean molecular weight species. We assume these species remain fully mixed in the envelope and do not consider processes such as condensation, rain-out or interior-envelope interactions. For a given set of model parameters, we utilise the planetary structure model from \citet{Rogers2025b} to determine a planet's transit radius. In the envelope structure model, we assume that water vapour is the dominant opacity carrier at the radiative-convective boundary and interpolate over the Rosseland mean opacity tables from \citet{Kempton2023} for H$_2$-He-H$_2$O mixtures. The structure model assumes an adiabatic equation of state for the convective envelope of the form $P \propto \rho^\gamma$, where $P$ and $\rho$ are gas pressures and densities, respectively. The adiabatic index, $\gamma$, is calculated from the mean degrees of freedom, given by:
\begin{equation}
    \bar{n}_\text{d} = x n_{\text{d,H}_2} + y n_{\text{d,He}} + z n_{\text{d,H}_2\text{O}}
\end{equation}
where $x$, $y$ and $z$ are mole fractions for H$_2$, He and H$_2$O, respectively, and $n_{\text{d,H}_2}=5$, $n_{\text{d,He}}=3$, $n_{\text{d,H}_2\text{O}}=6$ are the mean degrees of freedom for each species. Then the adiabatic index is calculated as $\gamma = (\bar{n}_\text{d} + 2) / \bar{n}_\text{d}$. 

The simplicity of our model is beneficial since it is agnostic to the planet's interior composition and structure. The only assumption is that a mass of $M_\text{c}$ sits below a radius of $R_\text{c}$. The specific composition, or state of matter within this interior is arbitrary and could represent any form, for example water-rich, water-poor, miscible or immiscible. We do not consider the full planetary evolution framework described in \citet{Rogers2025b}, but instead set the planet's cooling timescale equal to the host stellar age in order to prescribe the planet entropy as a proxy for thermal evolution \citep[see discussion from][]{Owen2020}. We label the planet's transit radius  and mass for a given set of model parameters as $R_\text{p,model}$ and $M_\text{p,model}$, respectively. 

The model parameters allow a hydrogen mass loss timescale to be calculated from Equation \ref{eq:MassLossTimescale}. The likelihood function is then given by:
\begin{equation}
\begin{split}
        \mathcal{L} & = \mathcal{G}(R_\text{p,model};\, R_\text{p,obs}, \sigma_{\text{R}_\text{p}}) \\ 
        & \times \mathcal{G}(M_\text{p,model}; M_\text{p,obs}, \sigma_{\text{M}_\text{p}}) \\
        & \times \mathcal{H}(\tau_{\dot{M}_\text{H}} - t_*),
\end{split}
\end{equation}
where the first two factors represent Gaussian functions, denoted as $\mathcal{G}(x; \mu, \sigma)$ for mean and standard deviation of $\mu$ and $\sigma$, that encapsulate the need for the model parameters to reproduce the planet's observed radius and mass. The final factor, $\mathcal{H}(\tau_{\dot{M}_\text{H}} - t_*)$, is a Heaviside step function that captures the inequality shown in Equation \ref{eq:X-inequality}:
\begin{equation}
    \mathcal{H}(\tau_{\dot{M}_\text{H}} - t_*) = \left\{
  \begin{array}{lr} 
      1 & \tau_{\dot{M}_\text{H}} \geq t_* \\
      0 & \tau_{\dot{M}_\text{H}} < t_*.
      \end{array}
\right.  
\end{equation}
In words, if the mass loss timescale of the modelled planet is shorter than the planet's lifetime, then the model is inconsistent with the presence of escaping hydrogen and the likelihood is zero. This encodes the lower limit on hydrogen mass described in Section \ref{sec:partI_sketch} into the inference model. We use the nested sampling algorithm of \verb|UltraNest| to determine posterior distributions with a minimum effective sample size of $5000$ and a minimum of $1000$ live points per epoch \citep{Buchner2021}.

\subsection{A note on priors: part I} \label{sec:priors_partI}
As with any Bayesian inference model, care must be taken when choosing priors. Here, we place uniform (flat) priors on the following model parameters: core mass, core radius, logarithm of $f_\text{env}$, and mean molecular weight. For other model parameters including semi-major axis, equilibrium temperature, host stellar mass and host stellar age, we place Gaussian priors based on their previously measured values as shown in Table \ref{tab:planets}. We place an additional prior on core mass and radius such that the core density cannot be lower than that of pure water ice or higher than that of pure iron. For this we use the mass-radius relations of \citet{Fortney2007}. 

Finally, for the observed planets of interest in this study (namely TOI-776 b, c and GJ-436 b), our priors on hydrogen mass loss rate come from posteriors inferred in \citet{Schreyer2024,Loyd2025}. These studies performed an inference analysis which involved fitting an escaping hydrogen tail model to \textit{HST} Ly-$\alpha$ transit observations in order to constrain properties such as planetary outflow velocity and mass loss rate. In the case of GJ-436 b, the relevant posterior from \citet{Schreyer2024} is simply a $1$D function of mass loss rate ($\log_{10} \dot{M}_\text{H}$). However, in the case of the TOI-776 planets, the posteriors on mass loss rate from \citet{Loyd2025} are conditional on planet mass. These $2$D functions form conditional priors within the \verb|UltraNest| framework. We note that the original analysis from \citet{Loyd2025} for TOI-776 b and c used the mass and radius values from \citet{Luque2021}. These values have since been revised in \citet{Fridlund2024}, which we adopt in this study. To do so, we resample the Monte Carlo Markov Chain (MCMC) posteriors from \citet{Loyd2025}, conditioned on the new planet mass measurements from \citet{Fridlund2024}. {\jr{This results in minor changes in the inferred mass loss rates, namely from $\log_{10} (\dot{M} / \text{g s}^{-1}) = 9.22^{+0.70}_{-0.83}$ to $\log_{10} (\dot{M} / \text{g s}^{-1}) = 9.25^{+0.63}_{-0.86}$ for planet b, and from $\log_{10} (\dot{M} / \text{g s}^{-1}) = 9.48^{+0.61}_{-0.71}$ to $\log_{10} (\dot{M} / \text{g s}^{-1}) = 9.18^{+0.56}_{-0.65}$ for planet c.}}

% For this particular inference problem, the correct parameterisation is a uniform prior on envelope hydrogen mass fraction, $X$, since our constraints come from the presence of a non-zero hydrogen mass loss rate which is (trivially) sensitive to the size of the planet's hydrogen mass reservoir. An alternative, but incorrect setup would be to re-parametrise the problem in terms of envelope mean molecular weight and apply a uniform prior on this variable. However, as shown in Figure \ref{fig:H2O-mmw-relation}, a strong non-linear relation exists between mean molecular weight and hydrogen mass fraction, meaning that a uniform prior on mean molecular weight would strongly bias results towards low hydrogen mass fractions, which is an inaccurate summary of our \textit{a priori} knowledge of each planetary system. Alternatively, a log-uniform prior on hydrogen mass fraction is also inappropriate since this unnecessarily biases the results to very low mean molecular weight.

\subsection{Results: constraints from mass loss}

\begin{figure*}
	\includegraphics[width=2.0\columnwidth]{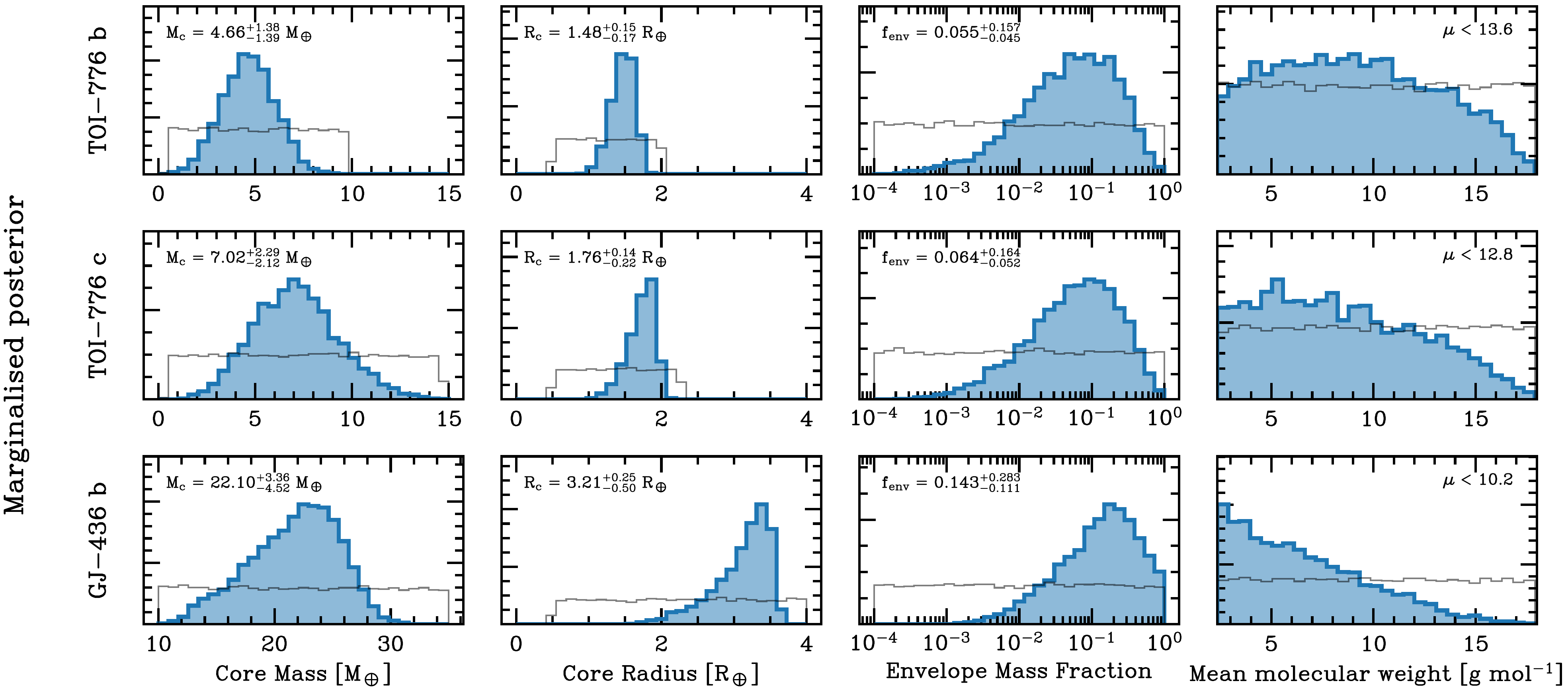}
    \centering
        \cprotect\caption{Marginalised posteriors are shown for the composition of TOI-776 b, TOI-776 c and GJ-436 b in the top, middle and bottom rows, respectively. The constraints come from observed masses, radii and escaping hydrogen mass loss rates from \citet{Schreyer2024,Loyd2025}. Here we show results for a selection of inferred parameters, including planetary core masses, $M_\text{c}$, core radii, $R_\text{c}$, envelope mass fractions $f_\text{env} \equiv M_\text{env} / M_\text{c}$, and envelope mean molecular weights $\mu$. Inferred values are quoted with their associated $1\sigma$ uncertainties (upper limits in the case of $1$-tailed posteriors). Grey histograms represent the prior probability distributions for each model parameter.} \label{fig:Structure_posteriors} 
\end{figure*} 

As a proof of concept, we now apply our Bayesian inference model to three observed sub-Neptunes that display confirmed hydrogen escape via Ly-$\alpha$ transits. We summarise the planetary and stellar parameters for our chosen planets, namely TOI-776 b, TOI-776 c, and GJ 436 b in Table \ref{tab:planets}. 

We present our marginalised posteriors for a selection of parameters in Figure \ref{fig:Structure_posteriors}, with results for TOI-776 b, TOI-776 c and GJ-436 b shown in top, middle and bottom rows, respectively. Note that prior distributions are shown in grey. For all planets, posteriors on core mass, core radius and envelope mass fraction are peaked with relatively wide distributions owing to the well-documented degeneracy in interior composition to explain a planet's mass and radius \citep[e.g.][]{Rogers2015,Rogers2023b}. For the largest planet, GJ-436 b, the posterior on mean molecular weight, $\mu$, is representative of a clear upper limit as predicted in Section \ref{sec:partI_sketch}. For a $1$-tailed distribution such as this posterior, the relevant $1\sigma$ upper limit equates to the $84^\text{th}$ percentile, which is $\mu < 10.2 \text{ g mol}^{-1}$ for GJ 436 b. This constraint is the strongest of the three planets in-part because the bulk density of GJ-436 b requires a low molecular weight compositional component, independently of any constraints from atmospheric escape. 

The smaller planets, TOI-776 b and c, are close to the upper edge of the radius valley, meaning that their interior composition cannot be constrained from just their mass and radius alone. Introducing the extra constraint from escaping hydrogen allows weak limits to be placed on atmospheric mean molecular weight. For TOI-776 b and c, we can place upper limits of $\mu < 13.6 \text{ g mol}^{-1}$ and $\mu < 12.8 \text{ g mol}^{-1}$, respectively. For these planets, particularly TOI-776 b, the posterior on mean molecular weight is marginally peaked at $\sim 7 \text{ g mol}^{-1}$. This is because the lowest values of mean molecular weight require smaller values of envelope mass fraction to fit the planet's radius. These hydrogen-rich envelopes have shorter mass loss timescales that are more likely to be shorter than the planet's age, thus down-weighting their likelihood. This is not true for larger mass planets such as GJ-436 b since they require massive hydrogen-rich envelopes with deep gravitational wells, thus precluding significant mass loss that can alter the planet's overall mass budget. The fact that a planet that sits within the radius valley, such as TOI-776 b, has an intermediate mean molecular weight is not surprising. Such planets are expected to be in the final stages of significant atmospheric loss as they transition to become super-Earths below the radius valley \citep{Loyd2025}. These planets may retain thin high mean molecular weight atmospheres, sculpted by mass loss and interior chemistry \citep[e.g.][]{Rogers2024b}.

% In Figure \ref{fig:LyA-MMW-posteriors}, we transform our marginalised posteriors to those of mean molecular weight. The strong non-linear relation between water mass fraction (or, more generally, ``metal'' mass fraction) and mean molecular weight asserts a higher probability at lower mean molecular weights. As a result, the posteriors for all planets present as upper limits, as expected from our simplified argument in Section \ref{sec:partI_sketch}. The $1\sigma$ upper limits on mean molecular weight for TOI-776 b, TOI-776 c and GJ-436 b are $5.75$~g mol$^{-1}$, $5.22$~g mol$^{-1}$ and $4.26$~g mol$^{-1}$, respectively. These equate to water mass fractions of \#, \# and \#, respectively.

% \begin{figure*}
% 	\includegraphics[width=2.0\columnwidth]{Figures/LyA_escape_mmw_posteriors.pdf}
%     \centering
%         \cprotect\caption{Marginalised posteriors for bulk envelope mean molecular weights for TOI-776 b, TOI-776 c and GJ-436 b in the left, middle and right-hand panels, respectively. Dashed lines show the one-tailed upper limits ($1\sigma = 84.1\%$, $2\sigma=97.7\%$). These upper limits are informed from observations of planet mass, radius and escaping hydrogen from \textit{HST}.} \label{fig:LyA-MMW-posteriors} 
% \end{figure*}

\section{Part II: Constraints from Atmospheric Spectroscopy} \label{sec:PartII}
The results from the previous analysis show that upper limits can be placed on atmospheric mean molecular weight for sub-Neptunes given observations of escaping hydrogen. However, this appears to be in contradiction to recent transit spectroscopy results, such as those from \textit{JWST}, for which multiple sub-Neptunes have been interpreted to host high mean molecular weight atmospheres via their relatively featureless spectra. In particular, the three planets considered in Section \ref{sec:partI} have all undergone atmospheric characterisation and have been interpreted to host either high mean molecular weight atmospheres and/or high altitude aerosols \citep{Alderson2025, Teske2025, Mukherjee2025}. In this section, we show that one can combine these analyses to infer joint posteriors on mean molecular weight, accounting for their escaping hydrogen atmospheres and featureless spectra. We perform this for TOI-776 c as a proof of concept, however, the methodology is applicable for any planet with observations of escaping hydrogen and atmospheric spectroscopy.

\subsection{Inference model for JWST spectra}
We first aim to reproduce the constraints on TOI-776 c's atmosphere from \textit{JWST} NIRSpec/G395H from \citet{Teske2025}. We make use of their \verb|Tiberius| data reduction \citep{Kirk2017,Kirk2021} and build isothermal atmosphere models using \verb|petitRADTRANS| \citep{Molliere2019}. As in \citet{Teske2025}, and in order for our models to be consistent with those in Section \ref{sec:partI}, we include H$_2$, He, and H$_2$O, including Rayleigh scattering and collision-induced absorption from H$_2$-He. For H$_2$O, we use the line list from \citet{Polyansky2018}. As in Section \ref{sec:partI}, we assume a Solar hydrogen to helium mass ratio of $X/Y = 0.7491 / 0.2377$ \citep[e.g.][]{Lodders2003} and are thus retrieving for the atmospheric water abundance that best fits TOI-776 c's spectrum. We also account for broadband opacity sources (such as clouds or aerosols) with an opaque pressure level. For each model, we use TOI-776's stellar radius of $R_* = 0.544 R_\odot$ and TOI-776 c's mass (see Table \ref{tab:planets}) to calculate the planet's transit depth, $\delta_\text{i}$, re-binned to match the $i^\text{th}$ wavelength bin in the \verb|Tiberius| data reduction from \citet{Teske2025}.

We broadly follow \citet{Teske2025} and include three physical parameters in our inference model: a parameterisation of metallicity (see discussion in Section \ref{sec:priors_partII}), opaque pressure level and reference pressure, corresponding to the pressure at which the planet's white-light radius and gravity are measured. We also include an NRS1/NRS2 detector offset parameter, measured in parts-per-million. The Gaussian likelihood is then given by:
\begin{equation}
    \mathcal{L} = \prod_i\mathcal{G}(\delta_i \,;\, \delta_{i,\text{obs}}, \sigma_i),
\end{equation}
where $i$ corresponds the $i^\text{th}$ wavelength bin and $\delta_{i,\text{obs}}$ are the observed transit depths with uncertainty, $\sigma_i$ from the \verb|Tiberius| data reduction from \citet{Teske2025}. As with the inference model in Section \ref{sec:partI}, we again use the nested sampling algorithm of \verb|UltraNest| to determine posterior distributions with a minimum effective sample size of $5000$ and a minimum of $1000$ live points per epoch \citep{Buchner2021}.

\subsection{A note on priors: part II} \label{sec:priors_partII}

For a relatively flat, featureless transmission spectrum, such as that of TOI-776 c, the choice of priors holds a subtle but important implication for interpreting atmospheric properties. Consider the opposite case of a spectrum that demonstrates clear molecular absorption bands. Here, ``free-retrievals'' are the current standard for inferring the presence of various gas species and thus mean molecular weights \citep[e.g.][]{Madhusudhan2009}. In this case, log-uniform priors are typically set on a species' mass, or molar abundance, owing to the fact that a very small abundance can still produce a significant opacity source and thus absorption feature. This is also the case for gas giants, in which one can confidently state that the atmosphere is hydrogen-dominated in order to explain the planet's mass and radius. In both of these cases, \textit{log-uniform priors on the abundance of a high mean molecular weight species are appropriate as long as the abundance of at least one of these species can be constrained.}

However, in our case of a flat spectrum, a retrieval will only be sensitive to relatively high abundances of high molecular weight gas species. This means that the likelihood function is uninformative at low gas abundance, resulting in a posterior distribution that remains unconstrained and simply follows that of the prior. {\jr{Since the results are sensitive to the choice in prior distribution, it is particularly important to make a choice that does not bias the results in an unphysical way.}} To intuit the best choice of prior, it is worth noting that a truly uninformative prior should be agnostic to the signal of the relevant experiment at hand. In our case, atmospheric spectroscopy signals are physically sensitive to the slant optical depth through a planet's atmosphere \citep[e.g.][]{Seager2000,Brown2001,Hubbard2001,Fortney2005,Miller-Ricci2009}. A planet's transit radius and transit depth as a function of wavelength scale with the scale height of the atmosphere \citep[see Equations 3 and 4 in][]{Kirk2025}. Since atmospheric scale height is inversely proportional to $\mu$, where $\mu$ is the gas mean molecular weight, then a truly uninformative prior distribution should be constant in $1 / \mu$. {\jr{We compare this choice of prior with other commonly adopted distributions in Appendix \ref{appendix:prior} and show that, unlike these commonly adopted choices, it does not bias the results in an unphysical manner.}}

For our proof-of-concept analysis, we follow a uniform prior in $1 / \mu$ alongside uniform priors for the NRS1/NRS2 detector offset, and log-uniform priors on opaque pressure level and reference pressure. We note that an analogous issue exists when attempting to infer the opaque pressure level since it, like atmospheric metallicity, also cannot be constrained in the case of a featureless spectrum. The combination of these effects means that the likelihood is uninformative for cloud pressures below $\sim 10^{-3} - 10^{-2}$~bar and the posterior in this region of parameter space reflects that of the joint prior placed on metallicity and cloud pressure. To avoid unnecessarily biasing the results to low mean molecular weight, we limit the log-uniform prior on opaque pressure levels between $10^{-5}$ and $1$~bar. This way, approximately half of the prior volume represents scenarios in which low-pressure aerosols mute all features, and the other half in which the featureless spectra can only be explained with high mean molecular weight. We encourage future theoretical studies on aerosols to also investigate the appropriate choices in Bayesian prior volume for grey opacity source parameterisation for such inference studies. 

% \begin{figure}
% 	\includegraphics[width=\columnwidth]{Figures/mmw_vs_H2O.pdf}
%     \centering
%         \cprotect\caption{The non-linear relation between water mass fraction and mean molecular weight in a H$_2$-He-H$_2$O gas mixture, assuming a Solar hydrogen-to-helium mass ratio, $X/Y = 0.7491 / 0.2377$.} \label{fig:H2O-mmw-relation} 
% \end{figure} 

\subsection{Results: constraints from atmospheric spectroscopy} \label{sec:results_partII}

\begin{figure*}
	\includegraphics[width=2.0\columnwidth]{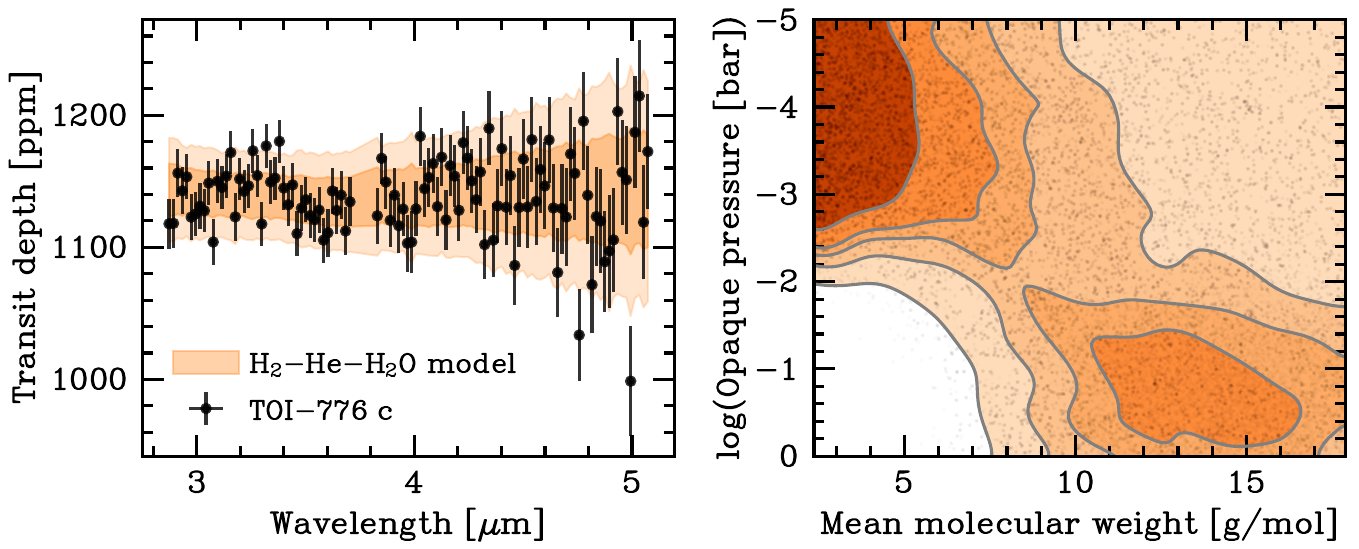}
    \centering
        \cprotect\caption{Left: \textit{JWST} NIRSpec/G395H spectrum for TOI-776 c from \citet{Teske2025} with our best-fit H$_2$-He-H$_2$O atmospheric model in orange. The shaded regions correspond to $1\sigma$ and $2\sigma$ confidence intervals. Right: marginalised posterior for atmospheric mean molecular weight and opaque pressure level. Contours represent increasing posterior probability (in $20\%$ intervals).} \label{fig:TOI-776c_JWST} 
\end{figure*} 

Our results are summarised in Figure \ref{fig:TOI-776c_JWST}. In the left-hand panel, we show the \verb|Tiberius| data reduction of the JWST NIRSpec/G395H transmission spectrum for TOI-776 c from \citet{Teske2025}. We also show the $1 \sigma$ and $2 \sigma$ bounds on our inferred best fit H$_2$-He-H$_2$O model as orange shaded regions. In the right-hand panel, we show the marginalised posterior as a function of atmospheric mean molecular weight and opaque pressure level. Here, contours represent increasing posterior probability density in increments of $20\%$ and black points represent individual samples from the \verb|UltraNest| chain. The posterior is bimodal. As also found in \citet{Teske2025}, at low aerosol pressures, all features are muted and so the posterior is only marginally updated from the prior, as further discussed in Appendix \ref{appendix:prior}. At aerosol pressures above $10^{-2}$~bar, the data requires high mean molecular weight to produce a small scale height and thus a flat spectrum. In this case, for opaque pressure $\lesssim 10^{-2}$~bar, we can rule out mean molecular weights $\lesssim 7 \text{ g mol}^{-1}$. 

\begin{figure*}
	\includegraphics[width=2.0\columnwidth]{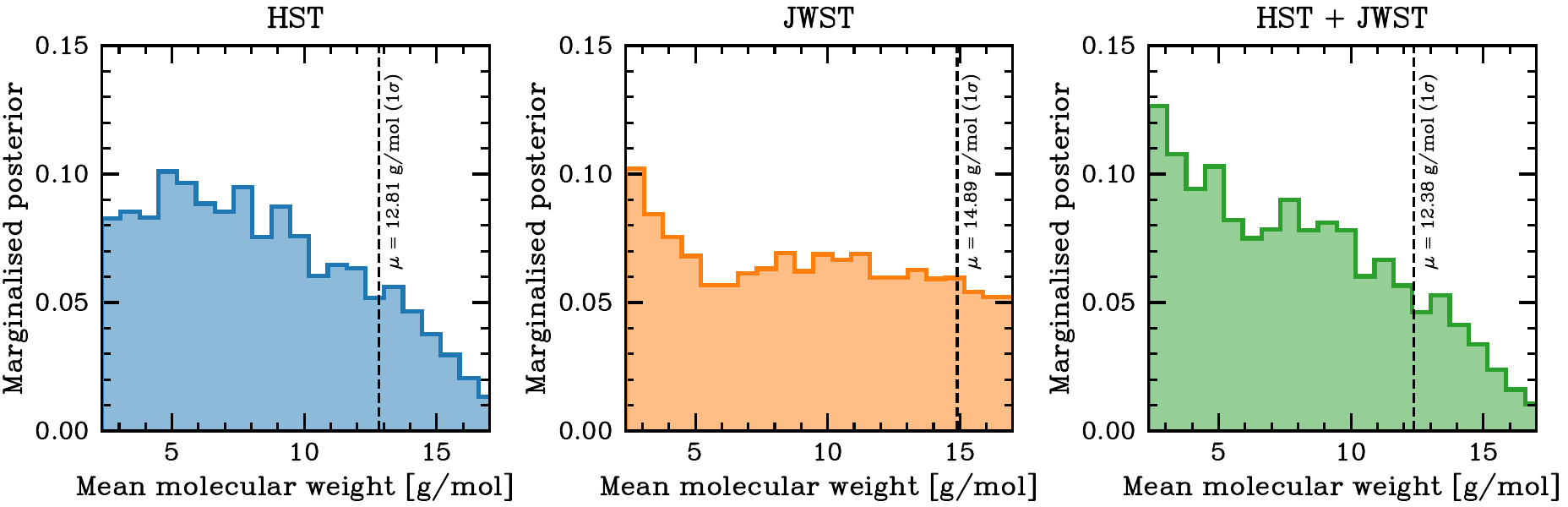}
    \centering
        \cprotect\caption{Marginalised posteriors for envelope mean molecular weights for TOI-776 c. In the left-hand panel we show the constraints from Section \ref{sec:partI}, namely Ly-$\alpha$ escaping exosphere observations from \textit{HST} (see Figure \ref{fig:Structure_posteriors}). In the central panel, we show the results from \textit{JWST} atmospheric transit spectroscopy from Section \ref{sec:PartII} (see Figure \ref{fig:TOI-776c_JWST}). In the right-hand panel, we combine these results to produce the joint posterior from \textit{HST} and \textit{JWST}.} \label{fig:HST+JWST_mmw_posteriors} 
\end{figure*} 

We marginalise across opaque pressure and show this posterior in the central panel of Figure \ref{fig:HST+JWST_mmw_posteriors} alongside the posterior for mean molecular weight from Section \ref{sec:partI} in the left-hand panel. The posterior on mean molecular weight from transmission spectroscopy presents as a weak upper limit of $\mu \leq 14.9 \text{ g mol}^{-1}$ ($1\sigma$). 

% We highlight that if the transformation from metal mass fraction to mean molecular weight is not performed, one could incorrectly interpret the posterior from Figure \ref{fig:TOI-776c_JWST} as a lower limit on mean molecular weight, as was done in \citet{Teske2025}, who claimed a lower limit of $\geq 6-8$~g mol$^{-1}$ to TOI-776 c. Similar conclusions were also made for TOI-776 b and TOI-836 b in \citet{Alderson2024, Alderson2025}.

\subsection{Combining constraints from mass loss and atmospheric spectroscopy} \label{sec:combinedResults}
The final part of our analysis is to combine the constraints found from the presence of escaping hydrogen (from Section \ref{sec:partI}) and atmospheric spectroscopy (from Section \ref{sec:results_partII}). We take the product of the posterior distributions on mean molecular weight from the two independent analyses, with the result shown in the right-hand panel of Figure \ref{fig:HST+JWST_mmw_posteriors}. {\jr{We find that atmospheric escape analysis significantly contributes to constraints on the composition and aerosol pressure for flat \textit{JWST} spectra.}} We can place a combined $1 \sigma$ upper limit on atmospheric mean molecular weight for TOI-776 c of $\mu \leq 12.4 \text{ g mol}^{-1}$. 

\begin{figure}
	\includegraphics[width=\columnwidth]{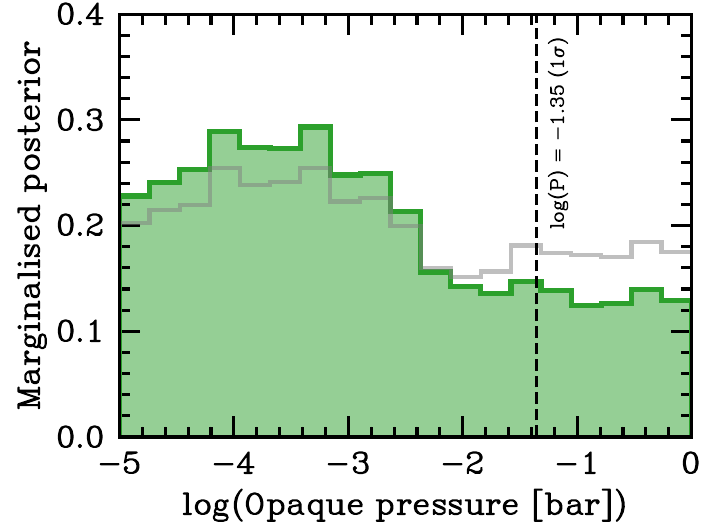}
    \centering
        \cprotect\caption{Marginalised posterior on opaque pressure level in the atmosphere of TOI-776 c. Constraints come from a combination of results from observed escaping hydrogen (Section \ref{sec:partI}) and transmission spectroscopy (Section \ref{sec:PartII}). The dashed black line highlights the $1\sigma$ upper limit of $P \leq 0.044$~bar. The grey histogram shows the marginalised posterior from transmission spectroscopy only.} \label{fig:Pcloud} 
\end{figure}

In Section \ref{sec:results_partII} we showed that the joint-posterior from \textit{JWST} transmission spectroscopy on mean molecular and opaque pressure level was bimodal. One peak corresponded to low mean molecular weight and low pressure aerosols, and the other to high molecular weight and high pressure aerosols. Combining these constraints with those from escaping hydrogen observations allows one to tentatively break this degeneracy, as shown in Figure \ref{fig:Pcloud}, in which the combined marginalised posterior for opaque pressure level is shown in green. The extra constraints from atmospheric escape update the posterior when compared to that of just transit spectroscopy (shown in grey) with preference for low pressure aerosols. If this is indeed the case for TOI-776 c and other similar planets, it suggests that such planets do indeed host low mean molecular weight atmospheres with spectral absorption features muted due to aerosols.

The constraint on mean molecular weight using this method can be significantly improved by reducing various measurement uncertainties. For TOI-776 c, the mass measurement uncertainty is currently $\sim 40\%$, the hydrogen escape rate uncertainty is $\sim 0.7$~dex, and the stellar age uncertainty is several billion years \citep{Fridlund2024, Loyd2025}. To illustrate how better constraints can be made, we construct a hypothetical scenario for a sub-Neptune with accurate mass, radius, mass loss and age characterisation, namely a radius of $R_\text{p}=3.0 \pm 0.1 R_\oplus$, mass of $M_\text{p}=8.0\pm0.8 M_\oplus$, hydrogen mass loss rate of $\log_{10}(\dot{M}_\text{H} / (\text{g s}^{-1})) = 9.5 \pm 0.1$, and host stellar age of $t_*=5.0+0.5$~Gyr. We then repeat the analysis from Section \ref{sec:partI}, assuming the same system parameters as TOI-776 c, e.g. semi-major axis, equilibrium temperature and host stellar mass. Our $1\sigma$ upper limit on mean molecular weight in this scenario is $\mu \leq 10.3 \text{ g mol}^{-1}$. If this posterior is combined with a spectrum matching that of TOI-776 c, the upper limit is further reduced to $\mu \leq 10.1 \text{ g mol}^{-1}$. We therefore encourage follow-up programs to better constrain these properties in order to better understand sub-Neptune composition.

\section{Discussion} \label{sec:discussion}
We have presented a new method of leveraging observations of escaping hydrogen to constrain the atmospheric mean molecular weight of sub-Neptunes. This method relies on a timescale argument in which the presence of escaping hydrogen is only possible if there is a sufficiently large hydrogen reservoir within the planet's envelope. In Section \ref{sec:PartII}, we also showed how these constraints can be combined with results from atmospheric characterisation, in this case transmission spectroscopy, to place joint constraints on atmospheric mean molecular weight. Here we discuss other applications of this method, including to observed helium escape, and specific structural constraints on temperate sub-Neptunes. 

\subsection{Sketch of applicability to helium escape}
Thus far we have only discussed using observations of escaping hydrogen via Ly-$\alpha$ transit spectroscopy to place constraints on atmospheric mean molecular weight. However, the use-case for the Ly-$\alpha$ feature is limited since it is only weakly sensitive to hydrogen mass loss rate \citep{Schreyer2024} and only probes neutral hydrogen, as opposed to ionised hydrogen for planets receiving higher X-ray/EUV flux \citep{Owen2023}. 

Another tracer of atmospheric escape exists via the helium $10830$~Å triplet absorption feature, which can be used as a probe of helium escape \citep[e.g.][]{Spake2018,Allart2018}. Since helium is also a low mean molecular weight gas species, one can reframe the analysis described in Section \ref{sec:partI} to place a lower limit on helium mass to explain prolonged escape at the observed rate (analogous to Equation \ref{eq:X-inequality}). The benefits of using helium escape is that the mid-infrared $10830$~Å feature is easier to observe, including from the ground, and does not suffer from the same interstellar medium extinction limitations as Ly-$\alpha$. Indeed, TOI-836 b has an observed escaping helium exosphere \citep{Zhang2025} and has also been observed in the \textit{JWST} COMPASS program to have a relatively featureless transmission spectrum \citep{Alderson2025}.

The greatest limitation, however, of using helium in our analysis stems from current theoretical shortcomings of linking the absorption feature to physical helium escape rates \citep[e.g.][]{Oklopcic2018,DosSantos2022,Ballabio2025,Schulik2025}. Unlike Ly-$\alpha$ absorption, it remains unclear exactly what the $10830$~Å feature actually probes and how to infer mass loss rates from its signal. Furthermore, mass fractionation induced by atmospheric escape may complicate the picture. In a hydrodynamic outflow, lighter species, such as hydrogen, can drag heavier species, such as helium, into the outflow \citep[e.g.][]{Zahnle1986,Hunten1987,Chassefiere1996}. In this case, the hydrogen-to-helium mass fraction remains constant with time. However, if the outflow velocity is insufficient to drag heavier species, then lighter species are preferentially lost and the mean molecular weight of the atmosphere will gradually increase with time \citep[e.g.][]{Cherubim2024}. Our theoretical understanding of this process is also currently limited since the collisional cross sections of gas species are strongly affected by factors such as outflow temperature and ionisation state \citep[e.g][]{Joselyn1978,Geiss1982}. Note that we have implicitly ignored mass fractionation in this proof-of-concept study by assuming hydrogen-to-helium mass fractions remain constant at Solar values.

With these caveats in mind, it is nevertheless worth sketching out how the inference method presented in Section \ref{sec:partI} could be adapted to account for helium escape and mass fractionation. From an observed helium escape rate, $\dot{M}_\text{He}$, one can calculate the hydrogen escape rate:
\begin{equation}
    \dot{M}_\text{H} = A \dot{M}_\text{He} \frac{X}{Y},
\end{equation}
where $X$ and $Y$ are envelope hydrogen and helium mass fractions and {\jr{$A\geq1$}} is an additional model parameter that accounts for mass fractionation. Then, one uses Equations \ref{eq:X-inequality} and \ref{eq:mmw-inequality} as before with the helium mass fraction now also being inferred. Of course an ideal scenario would be applying this methodology to a sub-Neptune for which escaping hydrogen \textit{and} helium have been observed, which would then allow constraints to be placed on mean molecular weight and mass fractionation processes. However, to date no such sub-Neptune has been observed \citep{DosSantos2023a}. 

Ongoing observational programs such as \textit{JWST} TUNES (The Unintentional NIRISS Escape Survey, AR-5916, PI: Vissapragada S.) and \textit{HST} STELa (Survey of Transiting Exoplanets in Lyman-alpha, GO-17804, PIs: Loyd R. O. P., Vissapragada S.) are well-suited to apply these methods to future observations of escaping hydrogen and helium. We highlight that improved accuracy on mass loss rates and planet mass allow for tighter constraints to be made on atmospheric mean molecular weight, as highlighted in Section \ref{sec:results_partII}.

\subsection{Sketch of applicability to testing the hycean world hypothesis}
Recent observations of temperate sub-Neptunes such as K2-18 b and TOI-270 d have sparked debate as to their interior compositions. Both of these planets have been hypothesised to be ``hycean worlds'', a theoretical class of water-rich sub-Neptunes in which a hydrogen-rich atmosphere is thin enough for water vapour to efficiently cool and condense out to form a liquid water ocean \citep[e.g.][]{Madhusudhan2021, Madhusudhan2020,Madhusudhan2023,Holmberg2024,Cooke2024,Rigby2024,Constantinou2025}. The atmospheric transmission spectra for these planets have revealed clear absorption features, unlike many of the hotter sub-Neptunes discussed in this study. For both of these planets, the non-detection of ammonia, NH$_3$, is contentious since it can be explained either by dissolution into a water ocean, or a magma ocean in the scenario that the planets comprise instead of hydrogen-rich envelopes atop Earth-like rocky interiors \citep[e.g.][]{Benneke2024,Wogan2024,Shorttle2024,Nixon2025}.

One of the clear predictions of the hycean hypothesis is that such planets should have thin hydrogen-dominated atmospheres in order for the liquid water ocean to be retained. For example, \citet{Rigby2024} predicted that in order for K2-18 b to be a hycean world, it requires a hydrogen/helium-dominated atmospheric mass fraction of $\lesssim 0.0052 \%$. Since the analysis we presented in Section \ref{sec:partI} can place constraints on hydrogen mass fraction, it is well-suited to testing the hycean hypothesis. Instead of inferring the mean molecular weight from an uninformative prior, one can use the well-constrained posterior from transmission spectra and infer the lower limit on envelope mass fraction, given some observations of escaping hydrogen or helium.

For K2-18 b, \citet{dosSantos2020} found tentative evidence for an escaping exosphere from \textit{HST} Ly-$\alpha$ observations. They observed a partial Ly-$\alpha$ transit of K2-18 b, finding a $\sim 70 \%$ decrease in the in-transit flux compared to the out-of transit flux in the blue-wing of line \citep[as seen in Figure 3 of][]{dosSantos2020}.  Since the data only consists of one partial transit, variation in the stellar Ly-$\alpha$ cannot be ruled out as the cause of this finding. Nevertheless, if we assume that the decrease in Ly-$\alpha$ flux is indeed due to hydrogen escaping from K2-18 b, we can estimate the hydrogen mass loss rate from K2-18 b, using a simple model of the geometry of a tail of escaping hydrogen from \citet{Owen2023}. The photoionisation timescale of a hydrogen atom in the outflow is estimated to be $\gtrsim 1000$ hours; therefore, the hydrogen in the tail can be assumed to be predominantly neutral. The minimum mass-loss rate required for the escaping tail to be optically thick is estimated by assuming the maximum Ly-$\alpha$ cross section of hydrogen at $10^4$ K. This is equivalent to assuming that the tail is accelerated radially outward such that the absorbing hydrogen is Doppler-shifted into the observable blue wing of the Ly-$\alpha$ line. We note that this represents the most conservative estimate of the mass-loss rate, since any deviation from this assumption would require a larger hydrogen column density to produce the same level of absorption. With this, we estimate a minimum hydrogen mass loss rate of $\sim 10^7 \text{ g s}^{-1}$, which is consistent with theoretically predicted energy-limited mass loss rates $\sim 10^8 \text{ g s}^{-1}$ \citep{dosSantos2020}. 

\begin{figure}
	\includegraphics[width=\columnwidth]{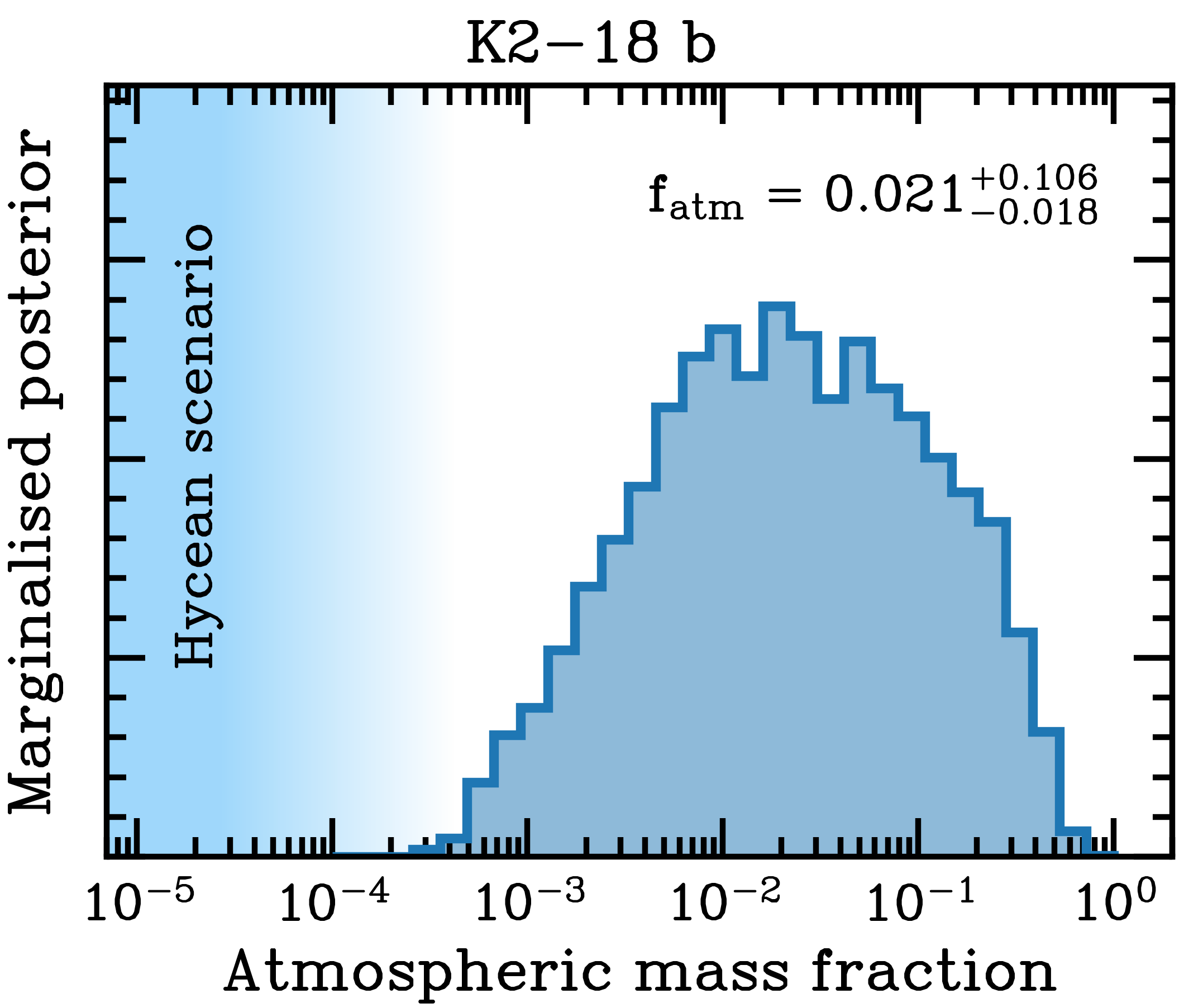}
    \centering
        \cprotect\caption{Marginalised posterior for the inferred atmospheric mass fraction of K2-18 b, given the observations of escaping hydrogen from \citet{dosSantos2020}. The blue shaded region denotes the approximate allowed range of atmospheric mass fractions for a hycean scenario, in which a thin hydrogen-rich atmosphere allows for the existence of a liquid water ocean \citep[e.g.][]{Rigby2024}} \label{fig:K2-18b} 
\end{figure}

As a rough proof-of-concept of this test on hycean worlds, we performed our analysis as described above for K2-18 b, with planetary and stellar parameters taken from \citet{Cloutier2019}, as well as a uniform prior on hydrogen mass loss rate from $7 \leq \log_{10}(\dot{M}_\text{H} / (\text{g s}^{-1})) \leq 8 $, in-line with the previous order-of-magnitude estimate, and a uniform prior on atmospheric mean molecular weight of $2.46 \leq \mu / \text{g mol}^{-1} \leq 7.64$ from the analysis of \citet{Schmidt2025}. As shown in Figure \ref{fig:K2-18b}, we constrain the atmospheric mass fraction to be $\log_{10} f_\text{env} = -1.67\pm0.78$, which is inconsistent with the upper limit from \citet{Rigby2024} of $\log_{10} f_\text{env} = -4.28$ at the $\sim 4 \sigma$ level. Again, we stress that the Ly-$\alpha$ observations from \citet{dosSantos2020} are tentative, and a rigorous analysis of converting such an observation to physical hydrogen mass loss rates has not been performed to date. Moreover, our model does not include hydrogen outgassing, which could in principle resupply a thin hydrogen atmosphere sourced from the interior (see discussion in Section \ref{sec:limitations}). Finally, the atmospheric mass fraction upper limit from \citet{Rigby2024} may vary with different choices in equation of state and pressure-temperature profiles. We recommend further observations of the escaping atmosphere from K2-18 b and other hycean candidates to test this hypothesis, as well as theoretical investigations of hycean evolution models which can determine the propensity for hydrogen outgassing.

\subsection{Model limitations and improvements} \label{sec:limitations}
The constraints on mean molecular weight presented in Section \ref{sec:partI} were dependent on an interior structure model. For this proof-of concept study, we chose a simple semi-analytic model from \citet{Rogers2025b}. One benefit of this model is that it is agnostic to the planet's interior composition, in that a planet's core mass and core radius are inferred as separate variables (meaning that the interior can be representative of any composition). However, the model is also limited with the assumption of ideal gases, polytropic equations of state and ignores chemical interactions between interior and envelope. Several studies have shown that chemical interactions can alter atmospheric composition \citep[e.g.][]{Chachan2018,Kite2020,kite2021,Schlichting2022,Misener2023}. Including these theoretical effects would couple planet properties such as core mass and radius with atmospheric mean molecular weight, which are assumed to be independant parameters in the current model. Furthermore, our model implicitly assumes that escaping hydrogen can only be sourced from the planet's envelope. Recent studies investigating the miscibility of silicate melt and hydrogen gas have shown that hydrogen may be stored in the interior and exsolved over time in a process analogous to outgassing, providing a larger reservoir for escaping hydrogen \citep{Young2024,Rogers2025c}. Miscibility has also been shown to affect the thermal evolution of sub-Neptunes, slowing contraction as a phase-change front moves radially inwards through the planet as silicate mass settles deeper into its gravitational well \citep{Rogers2025c}. Including these effects in the relevant parameter space may alter the inferred lower limits on hydrogen mass fraction presented in Section \ref{sec:partI}. In addition, another reservoir of hydrogen that we have ignored is within the atmospheric ``metals'' themselves. If photodissociated from molecular H$_2$O, hydrogen atoms could escape and be detected with Ly-$\alpha$, thus removing the requirement for another primordial hydrogen reservoir. However, for the high mass loss rates observed for the planets within this study ($\sim 10^9 \text{ g s}^{-1}$, see Table \ref{tab:planets}), the oxygen atoms from the dissociated molecular H$_2$O would also become entrained in the hydrogen outflow and thus reduce its sound speed, which is not observed \citep[see discussion in][]{Loyd2025}.

We recommend future studies to apply evolution models to this framework that are both flexible in terms of planet interior composition, but also self-consistent with regards to processes such as interior-envelope interactions \citep[e.g.][]{Young2024,Rogers2025c}, non-equilibrium processes such as photochemistry \citep[e.g.][]{Rigby2024,Nixon2025,Rigby2026}{\jr{, and metallicity-dependent atmospheric escape efficiencies \citep[e.g.][]{Yoshida2022,Broome2025}}}. Such modelling efforts will result in tighter constraints being made on atmospheric structure and composition due to regions of parameter space being ruled out on theoretical bounds. The combination of this framework with tighter measurement uncertainties on planet mass and atmospheric escape rates, as discussed in Section \ref{sec:combinedResults}, will allow for robust conclusions to be made for a given planet.

\section{Conclusions} \label{sec:conclusion}
The internal composition of sub-Neptunes is one of the most debated topics in exoplanetary science today. If indeed the leading hypothesis is correct and the small-planet radius valley was carved via atmospheric escape, then it requires sub-Neptunes to host low mean molecular weight envelopes \citep[e.g.][]{Owen2017,Bean2021,Rogers2025b}. However, recent atmospheric transit spectroscopy from \textit{JWST} has revealed several relatively flat spectra which are consistent with high mean molecular weight atmospheres \cite[e.g.][]{Wallack2024,Alderson2024,Mukherjee2025,Ahrer2025,Alderson2025,Teske2025}. However, many of these planets also have observed escaping hydrogen or helium exospheres \citep[e.g.][]{Ehrenreich2015,Zhang2025,Loyd2025,Ahrer2025}, pointing towards low-pressure aerosols being the cause of muted spectral features instead of high mean molecular weight atmospheres \citep[e.g.][]{Gao2021,Owen2025}.

In this study, we have presented a new technique that utilises observations of escaping hydrogen or helium to place upper limits on envelope mean molecular weight for a given planet. These constraints can be combined with those from atmospheric spectroscopy to better understand the atmospheric composition of sub-Neptunes. Our general conclusions are as follows:

\begin{itemize}
    \item A sub-Neptune requires a sufficiently large hydrogen or helium reservoir to supply sustained stellar-driven atmospheric escape. From this fact, one can place lower limits on envelope hydrogen or helium mass for a given observed escape rate and thus upper limits on envelope mean molecular weight.
    \vspace{0.4cm}

    \item From this method, we can place upper limits on the envelope mean molecular weight of TOI-776 b, TOI-776 c and GJ-436 b of $13.6$, $12.8$ and $10.2 \text{ g mol}^{-1}$, respectively. These results are indicative of such planets hosting low mean molecular weight, hydrogen-dominated atmospheres, however, such constraints can be significantly improved with better measurement uncertainty on planet mass, radius, atmospheric escape rate and planet age.
    \vspace{0.4cm}

    \item The inference on mean molecular weight from escaping atmosphere observations can be combined with constraints from atmospheric spectroscopy. We applied this to the relatively flat \textit{JWST} transit spectrum of TOI-776 c from \citet{Teske2025}. Although the mean molecular weight constraint is only marginally improved to $\mu \leq 12.4 \text{ g mol}^{-1}$, we can tentatively break the degeneracy between opaque pressure and atmospheric mean molecular weight, suggesting that the flat spectrum is indeed caused by high-altitude aerosols.
    \vspace{0.4cm}

    \item We have highlighted that care must be taken when choosing Bayesian priors in free-retrievals applied to relatively flat/featureless spectra of sub-Neptunes. We recommend priors that are uniform with respect to atmospheric scale height, or equivalently $\mu^{-1}$, so as to not unnecessarily bias the results to low mean molecular weight.
    \vspace{0.4cm}

    \item Finally, we have shown that our inference method can be reframed as a test of hycean worlds, a hypothesised class of water-rich sub-Neptunes with thin hydrogen-dominated atmospheres. We have applied this to the tentative detection of escaping hydrogen from K2-18 b from \citet{dosSantos2020}, and shown that if such a detection is robust, we can rule out the hycean scenario on K2-18 b at the $\sim 4 \sigma$ level. However, we stress that more observations are needed to confirm this result.

\end{itemize}

This study serves as a proof-of-concept that escaping atmospheres provide information as to the envelope composition of sub-Neptunes. We have highlighted that our results rely on interior structure evolution models, which warrant improvement in order to better constrain the composition of sub-Neptunes.

\section*{Acknowledgements}
{\jr{We kindly thank the anonymous reviewer for useful comments that helped improve the paper, as well as Johanna Teske, Natasha Batalha, Martin Binet, Shreyas Vissapragada, Parke Loyd, and Leonardo dos Santos for useful discussions that helped improve this paper.}} JGR gratefully acknowledges support from the Kavli Foundation. JEO is supported by a Royal Society University Research Fellowship. This project has received funding from the European Research Council (ERC) under the European Union’s Horizon 2020 research and innovation programme (Grant agreement No. 853022). JK acknowledges financial support from Imperial College London through an Imperial College Research Fellowship grant. ES acknowledges support from NASA XRP grants 80NSSC23K0282 and 80NSSC25K7153. This work is based in part on observations made with the NASA/ESA/CSA James Webb Space Telescope. The data were obtained from the Mikulski Archive for Space Telescopes at the Space Telescope Science Institute, which is operated by the Association of Universities for Research in Astronomy, Inc., under NASA contract NAS 5-03127 for JWST. These observations are associated with program \#$2512$.
%%%%%%%%%%%%%%%%%%%%%%%%%%%%%%%%%%%%%%%%%%%%%%%%%%
\section*{Data Availability}

All UltraNest chains are available at \href{https://zenodo.org/records/18269004}{10.5281/zenodo.18269004}. Models will be made available upon reasonable request.



\bibliographystyle{mnras}
\bibliography{references} % if your bibtex file is called example.bib

%\begin{thebibliography}{99}
%\end{thebibliography}

%%%%%%%%%%%%%%%%%%%%%%%%%%%%%%%%%%%%%%%%%%%%%%%%%%

%%%%%%%%%%%%%%%%% APPENDICES %%%%%%%%%%%%%%%%%%%%%

\appendix

\section{Choices in prior distribution for a flat transmission spectrum} \label{appendix:prior}

\begin{figure*}
	\includegraphics[width=2.0\columnwidth]{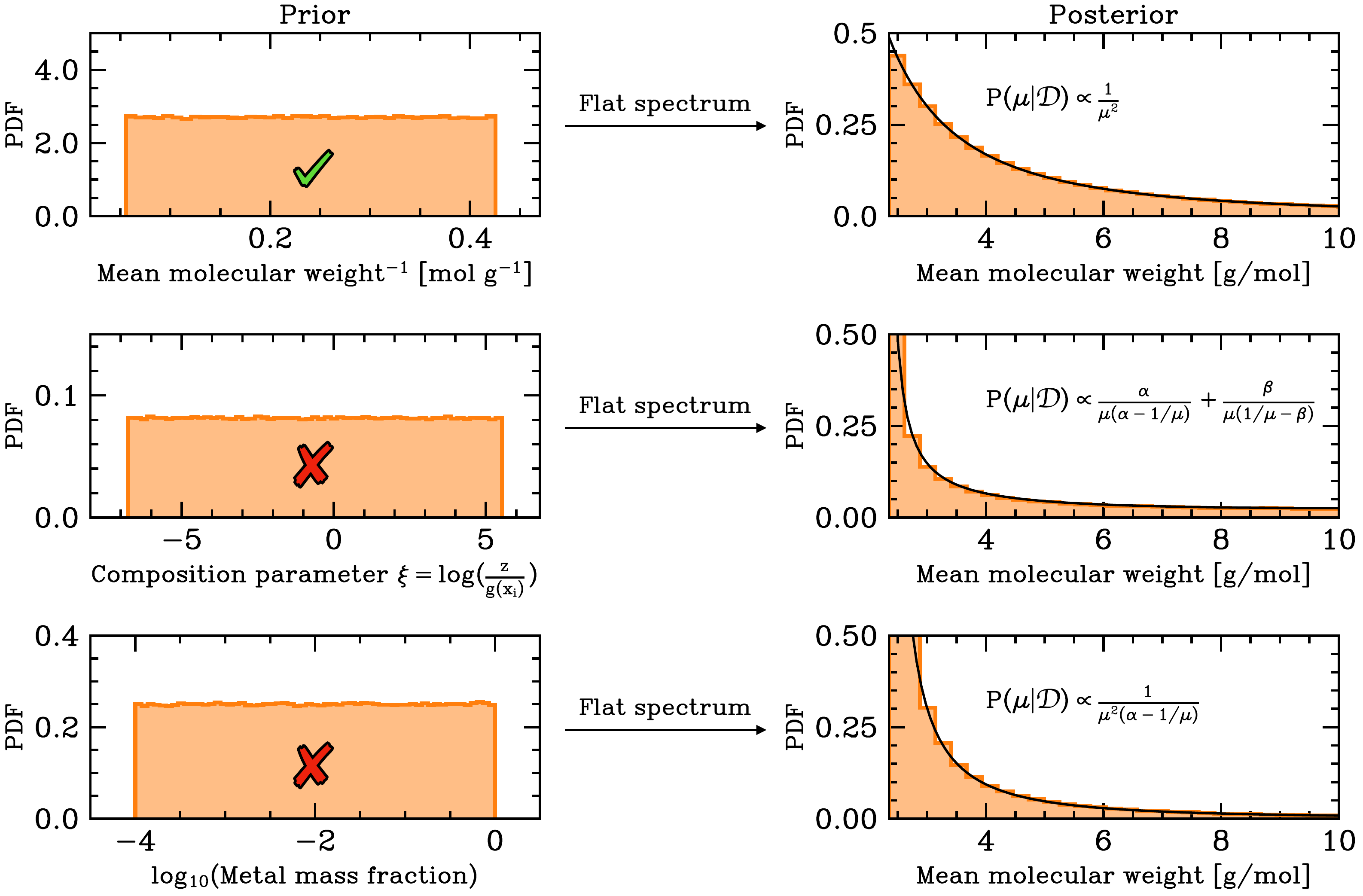}
    \centering
        \cprotect\caption{Three choices in prior parameterisation of atmospheric metallicity when applied to transmission spectroscopy in the case of a relatively featureless spectrum due to a high mean molecular weight atmosphere. If one wishes to place constraints on mean molecular weight, $\mu$, then we recommend a prior that is uniform with atmospheric scale height, or equivalently $\mu^{-1}$ (top row). Other options, including uniform in composition parameter \citep[middle row e.g.][]{Benneke2012} or log-scaled mass or molar abundance (bottom row) place an unnecessarily heavy posterior bias to low mean molecular weight. In the case of composition parameter, a heavy bias is also given to very high mean molecular weight, in this case at $\mu \sim 18$~g mol$^{-1}$, which is not visible in this figure.} \label{fig:prior_choice} 
\end{figure*} 

{\jr{In Section \ref{sec:priors_partII}, we have argued for an agnostic prior distribution that is constant in $1/\mu$, where $\mu$ is the mean molecular weight, when applied to a retrieval problem with a flat transmission spectrum.}} In Figure \ref{fig:prior_choice} we present a toy model of this problem. In the left-hand column we show various choices in prior distribution. We then assume a perfectly flat spectrum (with no noise), for example due to the presence of low pressure aerosols muting all spectral features and/or a high mean molecular weight. In this case, the likelihood is uninformative, and the posterior simply reflects that of the prior distribution. Since we are interested in the inferred atmospheric mean molecular weight, $\mu$, we transform the posterior to be a function of this variable via the relevant Jacobian:
\begin{equation}
    P(\mu | \mathcal{D}) = P(x(\mu)) \, \bigg|\frac{d x}{d \mu}  \bigg|,
\end{equation}
where $P(\mu | \mathcal{D})$ is the posterior probability density function (PDF) for data, $\mathcal{D}$, and $x(\mu)$ represents the chosen variable for the prior PDF, $P(x)$. The top row of Figure \ref{fig:prior_choice} shows our recommended choice in prior for a flat spectrum, which is constant as a function of atmospheric scale height, or equivalently constant in $x = \mu^{-1}$. In this case, one can show that for a prior with the form $P(\mu^{-1})=C_1$ between arbitrary limits, where $C_1$ is a constant, the posterior transforms to:
\begin{equation}
    P(\mu|\mathcal{D}) = \frac{C_1}{\mu^{2}},
\end{equation}
which is shown in the upper-right hand panel of Figure \ref{fig:prior_choice}. Note that this form naturally places a weak preference to low mean molecular weight owing to the requirement of agnosticism to the atmospheric scale height.

An alternative choice in prior, as shown in the second row of Figure \ref{fig:prior_choice}, is uniform in the composition parameter, $\xi$, defined as:
\begin{equation}
    \xi \equiv \ln \bigg(\frac{z}{g(x_i)}\bigg),
\end{equation}
where $z$ is the metal mole fraction and $g(x_i)$ is the geometric mean of all species mole fractions. In our simple parameterisation, the geometric mean is $(xyz)^{1/3}$, where $x$ and $y$ are the hydrogen and helium mole fractions, respectively. As first introduced within the context of atmospheric spectroscopy, \citet{Benneke2012} argued that this choice, referred to as the centred-log-ratio (CLR) prior, is the preferred option for atmospheres in which hydrogen cannot be assumed to be the dominant background gas. Its mathematical form ensures that all mole fractions sum to unity via uniform sampling from the relevant simplex and thus avoids preferentially biasing towards a high abundance of low mean molecular weight species \citep{Benneke2012}. In our simple case of a H$_2$-He-H$_2$O system, we again assume that mean molecular weight and gas species mass fractions are related via Equation \ref{eq:mmw-inequality}. For a fixed hydrogen-to-helium mass ratio, e.g. $\mathcal{R} \equiv X_\odot / Y_\odot$, we can then write:
\begin{equation} \label{eq:mmw-equality}
    z = \frac{\mu(1/\mu - \alpha)}{1 - \alpha/\beta},
\end{equation}
where $\alpha \equiv f_{\text{H}_2}/\mu_{\text{H}_2} + f_{\text{He}}/\mu_{\text{He}}$, $\beta \equiv 1 / \mu_Z$ with $f_{\text{H}_2} \equiv \mathcal{R}/(\mathcal{R}+1)$ and $f_{\text{He}} \equiv f_{\text{H}_2}/\mathcal{R}$. Then, for a CLR prior of the form $P(\xi)=C_2$, where $C_2$ is another constant, the posterior transforms as:
\begin{equation}
    P(\mu|\mathcal{D}) = \frac{2C_2}{3} \bigg( \frac{\alpha}{\mu(\alpha - 1/\mu)} + \frac{\beta}{\mu(1/\mu - \beta)} \bigg).
\end{equation}
Despite the convenience of the CLR prior in giving equal weight to each species in a permutation invariant manner, it strongly biases the posterior to both very low mean molecular weight and very high mean molecular weight, with the latter not visible in Figure \ref{fig:prior_choice}. This bias is much stronger than that of a prior that is constant in $\mu^{-1}$, as seen in Figure \ref{fig:prior_choice}. Although this bias softens as more species are added to the retrieval problem (which increases the dimensionality of the simplex), this is rarely a useful exercise for a featureless spectrum. We conclude, therefore, that the CLR prior is not agnostic to the transmission spectroscopy signal in this application.

% Note that this $1/\mu^2$ behaviour will change (to second order) if one relaxes some assumptions, for example if $\mu_Z$ is not constant due to chemical reactions at varying pressures and temperatures, or if the hydrogen-to-helium mass fraction, $\mathcal{R}$, does not remain at a fixed value, for example due to mass fractionation during atmospheric escape preferentially removing hydrogen instead of helium.

The final choice in prior distribution, which is shown in the bottom row of Figure \ref{fig:prior_choice}, is log-uniform in metal mass fraction, $Z$. Indeed this prior (or the molar mass equivalent) has been adopted in multiple studies on relatively flat sub-Neptune transmission spectra \citep{Ahrer2025,Wallack2024,Alderson2025,Teske2025,PelaezTorres2025}. In this case, for a prior of the form $P(\log_{10}Z)=C_3$, where $C_3$ is also a constant, the posterior transforms as:
\begin{equation}
    P(\mu|\mathcal{D}) = \frac{C_3}{\mu^2(\alpha - 1/\mu)}.
\end{equation}
 As shown in Figure \ref{fig:prior_choice}, this latter choice in prior distribution heavily biases the posterior to very low mean molecular weight and is therefore also unsuitable if one wishes to choose a truly agnostic prior for the transmission spectrum signal.

In summary, we recommend choosing a prior that is constant with atmospheric scale height, or equivalently constant in $\mu^{-1}$ for analyses of relatively featureless spectra. 

% \section{Some extra material}

% If you want to present additional material which would interrupt the flow of the main paper,
% it can be placed in an Appendix which appears after the list of references.

%%%%%%%%%%%%%%%%%%%%%%%%%%%%%%%%%%%%%%%%%%%%%%%%%%

% Don't change these lines
\bsp	% typesetting comment
\label{lastpage}
\end{document}